\documentstyle[11pt,paspconf,epsf]{article}

\begin{document}

\title{A Wide--Field Spectroscopic Survey In The Cluster Lens Cl0024+17}

\author{Oliver Czoske, Genevi\`eve Soucail, Jean--Paul Kneib}
\affil{Observatoire Midi--Pyr\'en\'ees, 14, Av.\ Ed.\ Belin,
  31400 Toulouse, France}

\author{Terry Bridges (AAO), Jean--Charles Cuillandre (CFHT), Yannick
  Mellier (IAP)}


\section{Introduction}

\begin{figure}[ht]
  \plotfiddle{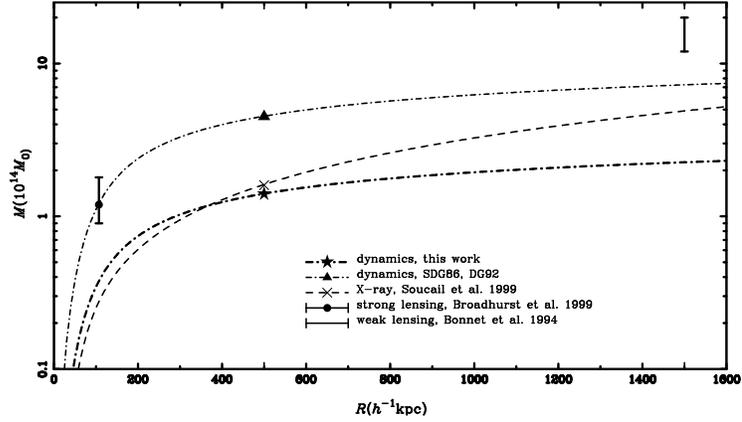}{48mm}{0}{80}{80}{-200}{-20}
  \caption{Compilation of mass estimates from dynamics, X--ray, strong
    and weak lensing} \label{discrep}
\end{figure}

In the past, studies of the lensing cluster Cl0024+17 have revealed a
strong discrepancy between estimates of the cluster mass using
different methods (galaxy dynamics, X--ray emission, strong and weak
lensing) of up to a factor of 3. These results are summarized in
Fig.\ \ref{discrep}. In order to better understand this discrepancy,
our group has undertaken a wide field spectroscopic survey in the
cluster.  

\section{Spectroscopic Sample}

 The largest redshift sample (prior to 1999) of members of
Cl0024 is the one by Dressler \& Gunn (1992), which included 31
redshifts resulting in a velocity dispersion $\sigma = 1200$ km/s. We
obtained a total of 626 spectra in the cluster field 
during three observing runs at CFHT in 1993, 1995 and 1996 and one
run at WHT in 1996. Adding to our catalogue the sample of 107 cluster
redshifts of Dressler et al.\ (1999) gives a total of 697
spectra. We obtain sufficiently secure redshifts for 615 objects in a 21 by
28 arcmin$^2$ field centred on the cluster.

\section{Results}

\begin{figure}[h]
  \plotfiddle{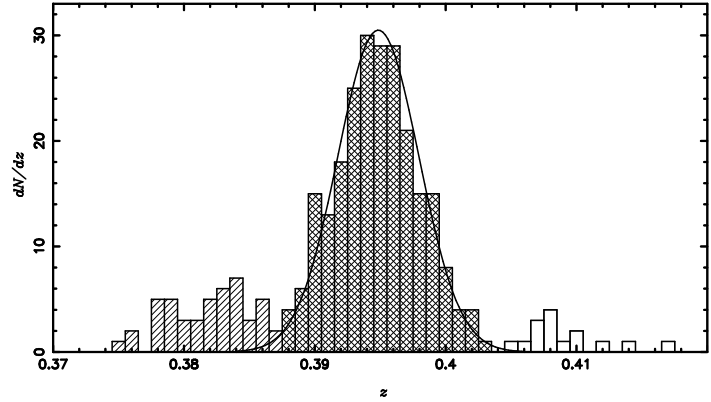}{49mm}{0}{80}{80}{-200}{-20}
  \caption{Velocity histogram around the cluster redshift showing a
    relaxed main cluster (cross--hatched) and an unrelaxed foreground
    extension (hatched). The curve is a Gaussian with our new velocity
    dispersion.} \label{histogram}
\end{figure}

The histogram (Fig.\ \ref{histogram}) shows the detailed structure of the velocity
distribution around the cluster redshift. We can clearly distinguish
the relaxed main cluster and an unrelaxed extension towards lower
redshifts. 

Defining as cluster members the 227 objects with redshifts
$0.388 < z < 0.405$, we find a central redshift of $\bar z =
0.3949\pm0.0006$ and a velocity dispersion $\sigma =
667^{+74}_{-51}\,\mbox{km/s}$ (biweight estimators, bootstrap
errors). A Gaussian with these parameters gives a good visual fit
(see Fig.\ \ref{histogram}). Using the simple spherically symmetric isothermal model of
Schneider, Dressler \& Gunn (1986), we obtain a mass for the main
cluster of $ M = 1.4 \times 10^{14}\,h^{-1}\, \mbox{M}_{\sun} $ (at 500~kpc).

We tentatively interpret the foreground extension as a filament 
aligned with the line of sight and estimate a lower limit for its mass
as $5\times10^{13} \,h^{-1}\, \mbox{M}_{\sun} $.
In order to better separate the contributions of the cluster and the
filament to the total lensing mass, a precise weak lensing mass
profile out to $\ga 1\,h^{-1}$~Mpc will be needed. 

\begin{references}
\reference Bonnet, H., Mellier, Y., Fort, B. 1994, \apj, 427, L83
\reference Broadhurst, T., astro-ph/9902316
\reference Dressler, A., Gunn, J. E. 1992, \apjs, 78, 1
\reference Dressler, A. et al.\ 1999, \apjs, 122, 51
\reference Schneider, D. P., Dressler, A., Gunn, J. E. 1986, \aj, 92, 523
\reference Soucail, G. et al.\ 1999, in preparation
\end{references}
\end{document}